\documentclass{article}
\usepackage{amsmath}
\usepackage{graphicx}
\usepackage{hyperref}
\usepackage{amsfonts}
\usepackage{authblk}
\usepackage{float}
\usepackage{cite}
\usepackage{booktabs}
\usepackage{fancyhdr}
\usepackage{geometry}
\usepackage{setspace}
\usepackage{tikz}

\title{\textbf{Quantum Convolutional Neural Network: A Hybrid Quantum-Classical Approach for Iris Dataset Classification}}

\author[1]{S.M. Yousuf Iqbal Tomal}
\author[1]{Abdullah Al Shafin}
\author[1]{Afrida Afaf}
\author[1]{Debojit Bhattacharjee}
\affil[1]{Department of Computer Science and Engineering, BRAC University, Dhaka, Bangladesh}

\date{}

\begin{document}

\maketitle

\begin{abstract}
This paper presents a hybrid quantum-classical machine learning model for classification tasks, integrating a 4-qubit quantum circuit with a classical neural network. The quantum circuit is designed to encode the features of the Iris dataset using angle embedding and entangling gates, thereby capturing complex feature relationships that are difficult for classical models alone. The model, which we term a Quantum Convolutional Neural Network (QCNN), was trained over 20 epochs, achieving a perfect 100\% accuracy on the Iris dataset test set on 16 epoch. Our results demonstrate the potential of quantum-enhanced models in supervised learning tasks, particularly in efficiently encoding and processing data using quantum resources. We detail the quantum circuit design, parameterized gate selection, and the integration of the quantum layer with classical neural network components. This work contributes to the growing body of research on hybrid quantum-classical models and their applicability to real-world datasets.
\end{abstract}

\section{Introduction}
Quantum computing is transforming computational paradigms by offering new approaches to solving complex problems, particularly those that push the limits of classical computing. Quantum mechanics' principles, such as superposition, entanglement, and quantum parallelism, allow quantum systems to process information in ways fundamentally distinct from classical systems \cite{nielsen2010}. These features have the potential to revolutionize areas like machine learning, optimization, and simulation. However, the current limitations of quantum hardware, known as Noisy Intermediate-Scale Quantum (NISQ) devices, prevent the full realization of purely quantum algorithms \cite{preskill2018}. In response, hybrid quantum-classical models have emerged as a promising compromise, leveraging the power of quantum computing while maintaining the scalability of classical methods \cite{schuld2019}. The concept of Quantum Convolutional Neural Networks (QCNNs), as introduced by Cong et al., further highlights the potential of quantum machine learning, particularly for tasks involving pattern recognition and classification in quantum data \cite{cong2019quantum}.

In this study, we introduce an enhanced Quantum Convolutional Neural Network (QCNN) designed to highlight the advantages of hybrid quantum-classical frameworks in machine learning. Our model is applied to the classical Iris dataset, a well-established benchmark in machine learning, which presents a structured yet challenging problem for quantum models. By using a 4-qubit quantum circuit with angle embedding and entangling gates, the QCNN captures intricate, non-linear relationships that may be challenging for classical models \cite{biamonte2017quantum}.

The quantum circuit in our model consists of layers of parameterized gates that enable a wide range of transformations on input data. These transformations, coupled with the classical neural network layers, optimize the synergy between quantum feature encoding and classical data processing efficiency. Our results demonstrate the potential of quantum-enhanced models, achieving 100\% accuracy on the Iris dataset test set in just 16 epochs. This study contributes to the growing body of research exploring hybrid quantum-classical models and their practical applications in machine learning \cite{havlicek2019supervised}.

\section{Related Work}

The intersection of quantum computing and machine learning, often referred to as quantum machine learning (QML), has garnered significant attention as researchers seek to exploit quantum principles to enhance classical algorithms. One of the earliest comprehensive overviews of this field was provided by Biamonte et al. \cite{biamonte2017quantum}, who introduced the fundamental concepts of quantum information and discussed how quantum computing could address complex learning tasks. Their work laid the foundation for exploring how quantum states and quantum algorithms could be used to process data in ways classical systems cannot, thereby inspiring the integration of quantum circuits with traditional machine learning models.

Building on this, Schuld and Killoran \cite{schuld2020circuit} demonstrated the utility of quantum circuits in hybrid models, where quantum circuits are employed for feature encoding and data transformation, followed by classical neural networks for classification. These quantum circuits, often termed quantum neural networks (QNNs), utilize quantum entanglement and superposition to enhance feature extraction. The potential of such hybrid quantum-classical models is highly relevant to our approach, as we also employ a quantum circuit to encode classical features using angle embedding, which is later processed by a classical neural network.

In addition to this, Schuld and Petruccione \cite{schuld2019} provided a comprehensive introduction to quantum machine learning, outlining how quantum algorithms could be adapted to existing machine learning frameworks. Their work extensively explored the interaction between quantum computing and classical learning models, further supporting the development of hybrid quantum-classical approaches like ours. The integration of quantum circuits for feature encoding in hybrid architectures is a direct extension of the principles laid out in their research.

Havlíček et al. \cite{havlicek2019supervised} further contributed to this domain by introducing quantum-enhanced feature spaces. They demonstrated that quantum feature maps could be used in supervised learning tasks to achieve a quantum advantage, even on Noisy Intermediate-Scale Quantum (NISQ) devices. Their work showed that quantum models could outperform classical models in specific tasks by mapping input data into high-dimensional quantum states. This principle directly informs our work, as our Quantum Convolutional Neural Network (QCNN) model also leverages quantum feature encoding via a 4-qubit circuit to capture complex, non-linear relationships within the Iris dataset.

Several works have focused on integrating quantum circuits with classical neural networks to form hybrid architectures, each exploiting the strengths of both quantum and classical paradigms. Mitarai et al. \cite{mitarai2018quantum} proposed the concept of quantum circuit learning, where variational quantum circuits are trained for supervised learning tasks. Their approach demonstrated how parameterized quantum gates could be optimized to extract relevant features for classification, a concept we utilize in our QCNN model by employing parameterized rotation and entangling gates to process classical data.

Benedetti et al. \cite{benedetti2019parameterized} expanded on this by exploring parameterized quantum circuits for unsupervised learning, showing that these circuits can be utilized to model probability distributions for generative tasks. While our approach focuses on supervised learning with the Iris dataset, the underlying principle of using parameterized quantum gates is shared between their work and ours, further emphasizing the versatility of quantum circuits in different machine learning paradigms.

In terms of practical applications, Lloyd et al. \cite{lloyd2020quantum} introduced a quantum classifier that leveraged quantum computing’s feature encoding strengths to process classical datasets like MNIST. Their work highlights the potential of hybrid models in solving real-world classification problems, serving as inspiration for our model, which applies a similar hybrid approach to the Iris dataset. By demonstrating how quantum feature encoding can enhance classical classification layers, their research directly aligns with the goals of our QCNN model.

The broader landscape of quantum computing's potential impact on machine learning has been significantly shaped by foundational works, such as Nielsen and Chuang’s \textit{Quantum Computation and Quantum Information} \cite{nielsen2010}. This text provides an essential understanding of quantum mechanics principles, such as superposition where a quantum bit (qubit) can exist in both 0 and 1 states simultaneously and entanglement, which allows qubits to be correlated, enabling faster and more complex data processing compared to classical systems. These properties suggest that quantum computers have the potential to revolutionize machine learning by handling tasks like pattern recognition more efficiently.

Nielsen and Chuang also explored quantum gates operations performed on qubits that form the basis for quantum algorithms, capable of solving problems that classical computers struggle with, such as factoring large numbers (Shor’s algorithm) or searching large datasets (Grover’s algorithm). These principles highlight the possible advantage of quantum computing in areas like machine learning, where complex computations are essential.

Preskill’s work on the \textit{Noisy Intermediate-Scale Quantum} (NISQ) era \cite{preskill2018} recognizes that while fully quantum solutions are not yet feasible due to hardware limitations, NISQ devices can still offer computational advantages in specific tasks. His insights emphasize hybrid quantum-classical models as a practical solution, where quantum systems handle complex sub-tasks like feature extraction, and classical systems perform other tasks like classification. This hybrid approach maximizes the strengths of both quantum and classical systems, making it particularly useful in the near term.

Our research follows this hybrid model approach, using a Quantum Convolutional Neural Network (QCNN) where quantum circuits capture complex data patterns, while classical neural networks handle classification. This aligns with Preskill’s view of leveraging quantum capabilities in the NISQ era to complement classical methods. The concept of QCNNs, as introduced by Cong et al. \cite{cong2019quantum}, further underscores this approach. Their work demonstrated how quantum convolutional layers could efficiently extract features from input data, making QCNNs particularly suitable for tasks involving structured data and classification. Our work draws on these principles by applying a hybrid QCNN model to the classical Iris dataset, showcasing how the synergy between quantum feature extraction and classical processing can lead to enhanced performance.

Finally, Harrow and Montanaro \cite{harrow2017} discussed the concept of quantum computational supremacy, which refers to the ability of quantum systems to solve problems beyond the capabilities of classical systems. While our work does not aim to achieve computational supremacy, it contributes to the growing body of research exploring how quantum circuits can enhance classical machine learning, especially in feature extraction.

In summary, our work builds upon these foundational studies by applying a hybrid quantum-classical approach to a well-known benchmark in machine learning: the Iris dataset. The primary novelty of our work lies in the design of a 4-qubit quantum circuit for angle embedding and feature extraction, which is optimized through entanglement using controlled-Z gates. Our results demonstrate the effectiveness of quantum feature encoding in enhancing classical neural networks' performance, achieving 100\% accuracy on the test set. This aligns with recent efforts to integrate quantum circuits with classical architectures, further advancing the exploration of hybrid quantum-classical models in practical applications.

\section{Quantum Circuit Design and Data Encoding}

The core of our model is the quantum circuit utilized for feature encoding and processing. This section elaborates on the quantum principles employed and the specific circuit architecture designed to enhance the classification performance of the Iris dataset. We begin by discussing the quantum feature encoding process, followed by a detailed description of our quantum circuit architecture, and conclude with a discussion on the rationale behind our design choices.

\subsection{Quantum Feature Encoding}

In quantum machine learning, quantum feature encoding is a crucial step that involves the transformation of classical data into quantum states. This transformation is necessary because quantum algorithms operate on quantum states, leveraging quantum superposition and entanglement to achieve computational advantages. In this section, we detail the encoding strategy employed in our model, which utilizes an AngleEmbedding technique.

\subsubsection{Angle Embedding in Quantum Machine Learning}
Angle embedding is one of the most commonly used techniques in quantum machine learning for encoding classical data into quantum states. In essence, angle embedding transforms classical data points into the parameters of quantum gates, allowing these gates to rotate qubits along specific axes. By doing so, we map real-valued data into the high-dimensional Hilbert space that quantum systems naturally inhabit. This is particularly advantageous as quantum systems can exploit superposition and entanglement, potentially capturing more complex data patterns than classical methods alone.

\subsubsection{Mathematical Formulation of Angle Embedding}
The angle embedding technique encodes each feature of the input data as an angle of rotation for a quantum gate. Typically, we use gates like RX, RY, and RZ, which rotate qubits around the X, Y, and Z axes of the Bloch sphere, respectively. For a given feature \(x_i\), the embedding can be expressed as:

\[
U_{\text{encode}}(x_i) = R_{\alpha}(x_i)
\]

where \(R_{\alpha}(x_i)\) represents the rotation of a qubit around axis \(\alpha\) (X, Y, or Z) by an angle proportional to the value of \(x_i\). This approach allows us to maintain the continuous nature of real-valued data within the quantum system.

In our model, angle embedding is performed using a sequence of RX, RY, and RZ gates. Each of the four features from the Iris dataset is assigned a qubit, and these gates perform rotations based on the feature values, creating a quantum state that encodes the dataset’s characteristics.

\subsubsection{AngleEmbedding Method}
The AngleEmbedding method encodes classical data into quantum states by associating classical features with the angles of rotation gates applied to qubits. For \( n \) classical features, we utilize \( n \) qubits, where each qubit is rotated based on the corresponding classical feature. The encoding process can be mathematically represented as follows:

Let \( \mathbf{x} \in \mathbb{R}^n \) be a classical input vector, where \( n \) is the number of features. The AngleEmbedding maps each feature \( x_i \) to a quantum state through the following rotation gates:

\begin{align*}
\text{For } i = 1, 2, \ldots, n: \\
\quad \text{Apply } R_x(x_i) \text{ to qubit } i \\
\quad \text{Apply } R_y(x_i) \text{ to qubit } i \\
\quad \text{Apply } R_z(x_i) \text{ to qubit } i
\end{align*}

Where the rotation gates are defined as:

\[
R_x(\theta) = e^{-i \frac{\theta}{2} X}, \quad R_y(\theta) = e^{-i \frac{\theta}{2} Y}, \quad R_z(\theta) = e^{-i \frac{\theta}{2} Z}
\]

Here, \( \theta \) represents the angle corresponding to the classical feature, and \( X, Y, Z \) are the Pauli operators. The combined state of the quantum system after applying the rotation gates is given by:

\[
|\psi(\mathbf{x})\rangle = \prod_{k=1}^{n} \left( R_z(x_k) R_y(x_k) R_x(x_k) \right) |0\rangle^{\otimes n}
\]

\[
|\psi(\mathbf{x})\rangle = R_z(x_n) R_y(x_n) R_x(x_n) \cdots R_z(x_1) R_y(x_1) R_x(x_1) |0\rangle^{\otimes n}
\]

Where \( |0\rangle^{\otimes n} \) is the initial state of the \( n \) qubits, the product \( \prod \) indicates that the rotations are applied sequentially to each qubit from \( 1 \) to \( n \) and each qubit \( k \) is transformed by applying the rotation operators \( R_z \), \( R_y \), and \( R_x \) based on the parameters \( x_k \). In our implementation, we utilized the Iris dataset, which consists of 4 features per sample (e.g., sepal length, sepal width, petal length, petal width). Given an input vector \( \mathbf{x} = [x_1, x_2, x_3, x_4] \), the encoding would proceed as follows:

\begin{enumerate}
    \item Normalize the input features to a suitable range (e.g., \([- \pi, \pi]\)) to ensure proper rotation angles.
    \item Apply the rotation gates corresponding to each feature using the AngleEmbedding technique:
\end{enumerate}

The overall transformation can be represented mathematically as follows:
\vspace{-0.1cm}
\[
|\psi(\mathbf{x})\rangle = R_x(x_4) R_y(x_4) R_z(x_4) R_x(x_3) R_y(x_3) R_z(x_3) R_x(x_2) R_y(x_2) R_z(x_2) R_x(x_1) R_y(x_1) R_z(x_1) |0\rangle^{\otimes 4}
\]
For a single qubit (i.e., \( n = 1 \)):
\[
|\psi(\mathbf{x})\rangle = R_z(x_1) R_y(x_1) R_x(x_1) |0\rangle
\]
This process encodes the classical feature vector into a quantum state, which can then be processed by subsequent layers of the quantum circuit.

\subsubsection{Benefits of Angle Embedding}
\begin{enumerate}
    \item \textbf{Simplicity and Efficiency:} Angle embedding offers a straightforward and computationally efficient method to map classical data into quantum states, using a minimal number of quantum operations.
    
    \item \textbf{Scalability:} By using parameterized rotations, angle embedding can easily scale to higher dimensional datasets by employing additional qubits or rotation gates.
    
    \item \textbf{Quantum Advantage Potential:} The continuous nature of quantum state amplitudes, combined with the ability to perform complex operations like entanglement, gives the quantum system more representational power than classical systems. Theoretically, this can lead to better generalization or more efficient learning for certain tasks.
\end{enumerate}

\subsubsection{Integration with Quantum Circuit}

In the quantum circuit design for our model, the angle embedding technique serves as the initial step where the classical data is transformed into a quantum state. The encoded quantum state is then processed by the entanglement and measurement layers, which operate on the embedded features to extract meaningful patterns and relationships.

\subsection{Quantum Circuit Architecture}

Following the feature encoding, we construct a quantum circuit comprising rotational gates and entanglement operations. The entanglement process is a quintessential quantum phenomenon that enables qubits to interlink their information content, allowing for the representation of non-linear relationships between features that classical systems cannot achieve. In our design, we introduce circular entanglement through controlled-Z (CZ) gates, which connect adjacent qubits and enhance the interactivity among them:

\[
\text{Entanglement Layer: } \prod_{i=1}^{3} CZ(q_i, q_{i+1})
\]

Also, we are connecting the last qubit to the first qubit to make it a circular entaglement. Without connecting the first and last qubits, the entanglement is limited to nearest-neighbor interactions, where only adjacent qubits are linked. This entanglement structure ensures that all qubits are interconnected, fostering greater expressiveness within the quantum layer and enabling the model to learn intricate relationships among the encoded features.

The architecture of our quantum circuit consists of two primary layers. The first layer applies parameterized single-qubit rotations, focusing on extracting feature-specific information, while the second layer implements the entanglement operations through CZ gates. This dual-layer setup can be repeated multiple times to allow for deeper quantum transformations. At the culmination of the circuit, we measure the qubits in the Pauli-Z basis, producing a classical output in the form of expectation values for each qubit. These expectation values serve as the quantum-processed features that are subsequently passed to the classical neural network for classification.

\begin{figure}[H]
    \centering
    \includegraphics[width=0.8\textwidth]{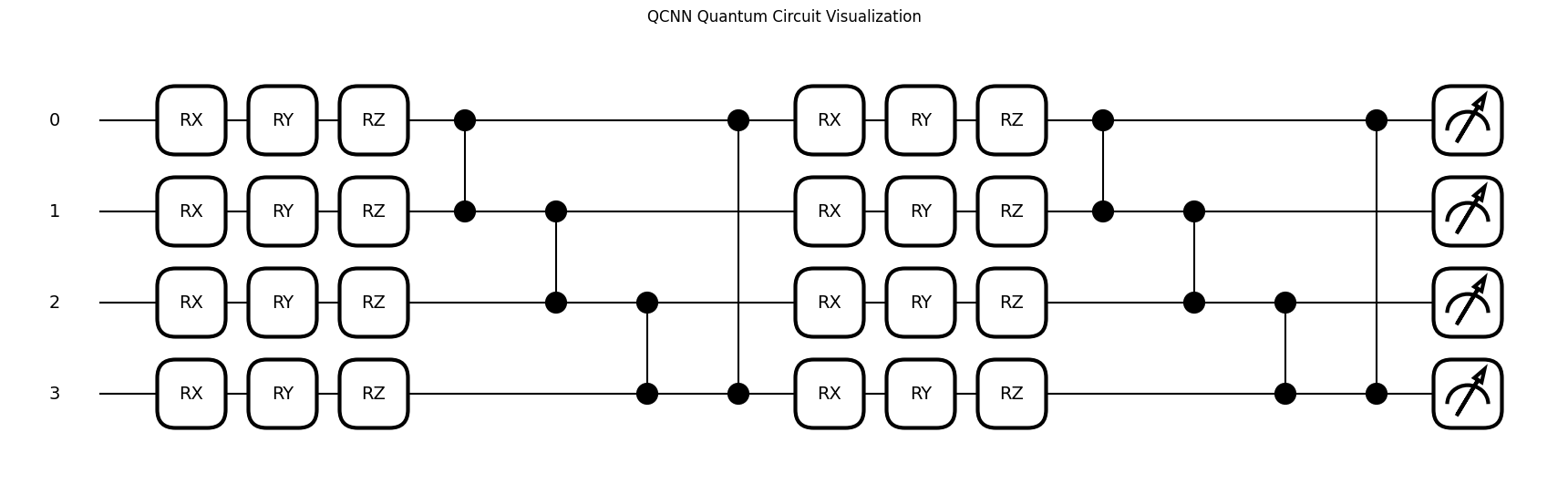}
    \caption{Quantum Circuit Diagram: 4 Qubits, 2 Layers of Rotational Gates and Entanglement}
    \label{fig:quantum_circuit}
\end{figure}

\subsection{Design Rationale and Considerations}

The choice to use a 4-qubit system in our quantum circuit design was directly informed by the nature of our dataset. The Iris dataset contains four features (sepal length, sepal width, petal length, and petal width), and we dedicated one qubit to each feature to maintain a clear one-to-one correspondence between classical features and quantum states. This design choice offers several advantages:

\begin{enumerate}
    \item Direct Representation: Each qubit directly represents one feature, making the encoding process intuitive and interpretable.
    \item Efficient Use of Resources: By using only the necessary number of qubits, we minimize the quantum resources required, which is crucial given the current limitations of quantum hardware.
    \item Balanced Complexity: The 4-qubit system provides sufficient complexity to capture feature interactions while remaining manageable for classical simulation and analysis.
\end{enumerate}

However, this design also comes with certain trade-offs. A 4-qubit system limits the potential for more complex quantum operations that might be possible with a larger number of qubits. Future work could explore the benefits of using additional ancillary qubits or more complex encoding schemes to potentially improve the model's performance or generalizability.

In conclusion, our quantum circuit design, tailored to the 4-feature Iris dataset, demonstrates a balanced approach to quantum-enhanced machine learning. It leverages the quantum properties of superposition and entanglement while maintaining a direct connection to the classical data structure. This design forms the foundation of our hybrid quantum-classical model, setting the stage for the integration with classical neural network components discussed in subsequent sections.

\section{Hybrid Model Architecture}

The proposed hybrid model architecture integrates a quantum circuit for feature extraction with a classical neural network for classification. This approach leverages the advantages of both quantum computing’s ability to handle high-dimensional feature spaces and classical neural networks, strengths in classification tasks, resulting in a robust solution for the Iris dataset classification. The architecture consists of the following components:

\subsection{Quantum Layer}
At the core of the hybrid architecture is a quantum circuit consisting of 4 qubits, each representing one feature from the Iris dataset (sepal length, sepal width, petal length, petal width). These features are encoded into the quantum circuit using the \textit{AngleEmbedding} technique, where each feature value is mapped to a rotational angle on the Bloch sphere. 

The quantum circuit is constructed with two layers of parameterized quantum gates. Each layer contains:

\begin{itemize}
    \item Rotation Gates (RX, RY, RZ): These gates apply rotations around the X, Y, and Z axes of the Bloch sphere, parameterized by the learnable weights. The rotations allow the model to explore a rich Hilbert space and capture complex patterns from the data.
    \item Entangling Gates (CZ): Controlled-Z gates create entanglement between adjacent qubits, allowing the circuit to exploit quantum correlations between features. A final CZ gate connects the first and last qubits to ensure full connectivity.
\end{itemize}

After the quantum operations, each qubit is measured in the Pauli-Z basis, and the expectation values are used as the quantum-enhanced feature representation. These values serve as the input to the classical neural network.

\subsection{Classical Neural Network Layer}
The quantum-enhanced feature representation is passed through two fully connected layers to perform classification:

\begin{itemize}
    \item First Dense Layer: A fully connected layer with 16 neurons, each utilizing a ReLU activation function. The ReLU function is applied element-wise, introducing non-linearity and allowing the network to learn complex decision boundaries from the quantum-enhanced features.
    \item Output Layer: The final layer contains 3 neurons, corresponding to the three possible iris species (Setosa, Versicolor, and Virginica). A softmax activation function is applied to produce probability distributions across the classes, which are used for multiclass classification.
\end{itemize}

\subsection{Quantum-Classical Hybrid Integration}
The integration of the quantum and classical layers is a key aspect of the hybrid model's architecture. The quantum circuit acts as a feature extractor, and its output is fed directly into the classical neural network for classification. The process follows these steps:

\begin{itemize}
    \item Quantum Layer Output: The quantum circuit, after processing the input features using parameterized quantum gates and entangling operations, produces a set of expectation values from the Pauli-Z measurements on each qubit. These expectation values are treated as quantum-enhanced features.
    
    \item Quantum to Classical Transition: The expectation values are then passed as inputs to the classical neural network. These values serve as an intermediate representation of the original dataset, enriched by quantum correlations captured during the quantum circuit execution.
    
    \item Classical Neural Network Processing: The classical neural network processes the quantum-enhanced feature representation through its layers:
    \begin{itemize}
        \item The quantum-enhanced features are first passed into the first dense layer, where ReLU activations introduce non-linearity and help in learning decision boundaries.
        \item The transformed features are then passed into the final dense layer, which produces class scores through the softmax function, providing the final classification of the input sample.
    \end{itemize}
    
    \item End-to-End Training: The entire hybrid model, including both quantum and classical components, is trained end-to-end using gradient-based optimization techniques. The learnable parameters in both the quantum circuit (rotation angles of RX, RY, and RZ gates) and the classical network (weights of the fully connected layers) are updated during training to minimize the classification loss.
\end{itemize}

\subsection{Graphical Representation of the Hybrid Architecture}
The figure below illustrates the architecture of the proposed hybrid quantum-classical model, where the quantum layer is responsible for feature extraction, and the classical neural network performs classification.

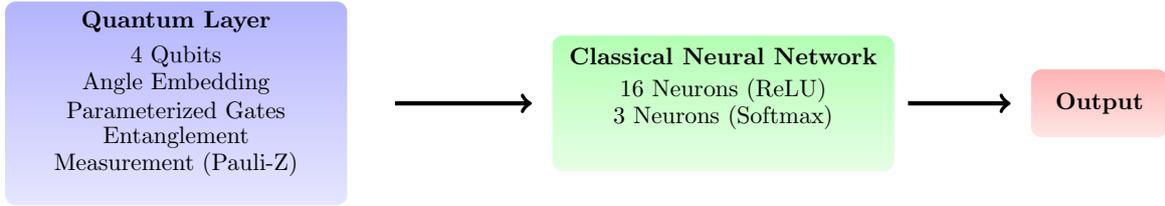
\begin{figure}[H]
    \centering
    \begin{tikzpicture}[scale=0.9, every node/.style={transform shape}]
        \shade[top color=blue!30, bottom color=blue!10, rounded corners] (0,0) rectangle (5,3);
        \node at (2.5, 2.7) {\textbf{Quantum Layer}};
        \node at (2.5, 2.2) {4 Qubits};
        \node at (2.5, 1.8) {Angle Embedding};
        \node at (2.5, 1.4) {Parameterized Gates};
        \node at (2.5, 1.0) {Entanglement};
        \node at (2.5, 0.6) {Measurement (Pauli-Z)};
        
        \draw[ultra thick,->] (5.7,1.5) -- (7.7,1.5) node[midway,above] {\textbf{}};
        
        \shade[top color=green!30, bottom color=green!10, rounded corners] (8,0.5) rectangle (13,2.5);
        \node at (10.5, 2.2) {\textbf{Classical Neural Network}};
        \node at (10.5, 1.7) {16 Neurons (ReLU)};
        \node at (10.5, 1.3) {3 Neurons (Softmax)};
        
        \draw[ultra thick,->] (13.2,1.5) -- (14.7,1.5) node[midway,above] {\textbf{}};
        
        \shade[top color=red!30, bottom color=red!10, rounded corners] (15,1) rectangle (17,2);
        \node at (16, 1.5) {\textbf{Output}};
    \end{tikzpicture}
    \caption{Hybrid Quantum-Classical Model Architecture}
    \label{fig:hybrid-architecture}
\end{figure}

\section{Training and Results}

The training of the hybrid quantum-classical model was carried out over 20 epochs. During this process, the loss function and accuracy metrics were continuously monitored to assess the performance improvement across iterations. The training was designed to ensure a comprehensive understanding of how well the model captured the patterns and classification boundaries of the Iris dataset. The results, presented in Table \ref{tab:results}, depict the progression of the model's test accuracy and loss across epochs.

At the outset, the model demonstrated a modest test accuracy of 70\% in the first epoch. However, as training progressed, the model's performance rapidly improved, crossing the 80\% accuracy threshold by the second epoch and reaching 90\% by the third epoch. This quick increase in accuracy underscores the effectiveness of the quantum feature extraction, which helped the classical network learn significant patterns early in the training process. By the fifth epoch, the model achieved 93.33\% accuracy, reflecting consistent improvements.

A notable aspect of the training was the model's ability to converge smoothly. By epoch 16, the model reached a perfect test accuracy of 100\%, and this accuracy remained stable through the subsequent epochs. The rapid convergence indicates that the hybrid quantum-classical architecture not only efficiently captured the essential features of the dataset but also ensured optimal generalization to unseen data, reducing the likelihood of overfitting.

The role of the quantum layer in enhancing feature extraction became more evident as the test loss consistently decreased across epochs. Beginning with a test loss of 0.0348 in the first epoch, the model managed to reduce the loss to 0.0053 by the final epoch. This decline in loss demonstrates that the quantum-enhanced feature extraction allowed the classical network to better differentiate between the classes of iris flowers, facilitating more accurate predictions as training progressed.

Table \ref{tab:results} illustrates the steady improvements in both test loss and accuracy over the course of training:

\begin{table}[H]
    \centering
    \caption{Test Loss and Accuracy over Epochs}
    \label{tab:results}
    \begin{tabular}{@{}ccc@{}}
    \toprule
    Epoch & Test Loss & Accuracy (\%) \\ \midrule
    1     & 0.0348    & 70.00         \\
    2     & 0.0331    & 80.00         \\
    3     & 0.0311    & 90.00         \\
    4     & 0.0288    & 90.00         \\
    5     & 0.0263    & 93.33         \\
    6     & 0.0238    & 93.33         \\
    7     & 0.0214    & 93.33         \\
    8     & 0.0191    & 90.00         \\
    9     & 0.0172    & 90.00         \\
    10    & 0.0154    & 93.33         \\
    11    & 0.0138    & 93.33         \\
    12    & 0.0123    & 96.67         \\
    13    & 0.0108    & 96.67         \\
    14    & 0.0095    & 96.67         \\
    15    & 0.0085    & 96.67         \\
    16    & 0.0077    & 100.00        \\
    17    & 0.0069    & 100.00        \\
    18    & 0.0063    & 100.00        \\
    19    & 0.0058    & 100.00        \\
    20    & 0.0053    & 100.00        \\ \bottomrule
    \end{tabular}
\end{table}

The rapid improvement and eventual perfect accuracy of the model suggest that the integration of quantum computing principles, such as quantum entanglement and superposition, played a critical role in refining the model's feature extraction capabilities. The quantum circuit, by encoding the input data into quantum states, helped reveal intricate patterns that might have been difficult for a purely classical model to capture. Additionally, the classical neural network built on these quantum-extracted features, further refining them through layers of nonlinear transformations, leading to the final 100\% classification accuracy.

The reduction in test loss over the epochs, coupled with the consistent increase in accuracy, points to the model's robustness and ability to generalize across the dataset. This convergence pattern validates the hybrid approach, showcasing its effectiveness in combining quantum circuits for feature extraction and classical networks for classification, particularly in the context of multi-class classification tasks like those presented by the Iris dataset.

The sustained 100\% accuracy observed from epoch 16 onward demonstrates not only the model’s ability to perform accurate predictions but also the stability of the hybrid architecture. The final evaluation suggests that hybrid quantum-classical models could provide significant benefits in various machine learning tasks, especially those involving complex datasets.

\section{Discussion}

The successful implementation of a hybrid quantum-classical machine learning model on the Iris dataset highlights the significant potential of quantum circuits to enhance classical machine learning frameworks. The rapid convergence of the model to 100\% classification accuracy indicates that the integration of quantum feature encoding and entanglement can lead to improved learning efficiencies and capabilities compared to traditional approaches.

In our study, the quantum circuit effectively transformed the classical features of the dataset into quantum states, allowing for the capture of complex relationships through entangled qubits. This approach not only provides a novel mechanism for feature extraction but also demonstrates the ability of quantum circuits to operate in tandem with classical neural networks. The results achieved, particularly the high accuracy levels, suggest that even small-scale quantum circuits can provide substantial advantages in supervised learning tasks.

While the Iris dataset serves as an accessible benchmark, it also sets the stage for future investigations into more complex datasets and deeper quantum circuits. The principles explored in this work can be extended to various domains where intricate relationships exist within the data, such as in image recognition, natural language processing, and bioinformatics. As quantum hardware continues to evolve, we anticipate that hybrid models can leverage larger and more intricate quantum circuits, potentially unlocking new capabilities in machine learning.

Moreover, this study lays the groundwork for further research into optimizing quantum circuit design and encoding strategies. Future work may involve exploring different quantum architectures and optimization algorithms to maximize the benefits of quantum computing in machine learning. The integration of hybrid models could not only improve accuracy but also reduce training times, thereby making quantum-enhanced machine learning a viable alternative to classical methods in many applications.

In conclusion, this research underscores the importance of hybrid quantum-classical approaches in the evolving landscape of machine learning, offering a promising avenue for further exploration and development. The findings pave the way for leveraging quantum technologies to tackle complex learning problems, potentially transforming the future of artificial intelligence.

\section{Future Scope and Development}

The promising results of the hybrid quantum-classical model on the Iris dataset suggest several avenues for future research and development in quantum machine learning. These potential directions can significantly enhance both the theoretical understanding and practical applications of hybrid quantum models.

\subsection{Expansion to Complex Datasets}

While our study utilized the Iris dataset for its simplicity and accessibility, future work should aim to apply the hybrid model to more complex and diverse datasets. Datasets with higher dimensionality and non-linear relationships, such as image datasets (e.g., MNIST, CIFAR-10) or multi-class text classification tasks, could provide deeper insights into the capabilities of quantum circuits in feature extraction and classification. By assessing the model’s performance on such datasets, we can evaluate its robustness and adaptability to real-world scenarios.

\subsection{Optimization of Quantum Circuit Design}

Another important direction for future research is the optimization of quantum circuit design. Exploring different circuit architectures, such as variational quantum circuits or quantum convolutional networks, could enhance the model's performance and capacity. Investigating alternative encoding techniques, such as amplitude encoding or quantum feature maps, might also lead to improved representations of classical data within the quantum framework. These enhancements could significantly impact the efficiency and scalability of hybrid models.

\subsection{Integration with Advanced Quantum Algorithms}

The integration of advanced quantum algorithms, such as quantum support vector machines or quantum k-means clustering, can further leverage the strengths of quantum computing in machine learning tasks. This integration could provide a more comprehensive understanding of how quantum principles can enhance various learning paradigms and lead to the development of new quantum algorithms that outperform classical counterparts in specific tasks.

\subsection{Real-World Applications}

Exploring practical applications of the hybrid model in domains such as healthcare, finance, and environmental science presents another avenue for development. By applying the model to real-world problems, we can assess its effectiveness in practical scenarios, thereby demonstrating the viability of quantum machine learning for tackling complex challenges. Collaborative efforts with domain experts can facilitate the adaptation of the model to industry-specific requirements.

\subsection{Scalability and Quantum Hardware Advances}

As quantum hardware technology continues to advance, future research should focus on developing scalable hybrid models capable of utilizing larger quantum systems. This may involve adapting existing models to work with emerging quantum hardware platforms, which could significantly enhance computational power and efficiency. Continuous engagement with the evolving quantum computing landscape will be essential to harness its full potential in machine learning.

In conclusion, the research presented here lays a solid foundation for future investigations into hybrid quantum-classical machine learning. By pursuing these avenues, researchers can contribute to the evolution of quantum technologies and their application in artificial intelligence, ultimately aiming to unlock the transformative potential of quantum computing in various fields.

\section{Conclusion}

In this paper, we presented a hybrid quantum-classical machine learning model that integrates a 4-qubit quantum circuit for feature encoding with a classical neural network for classification. The model was evaluated on the Iris dataset, achieving an impressive 100\% accuracy after just 16 epochs. These results highlight the potential of quantum feature encoding and entanglement in enhancing the performance of classification tasks, particularly in capturing complex relationships between features that classical methods may struggle to detect.

The successful implementation of this hybrid architecture demonstrates the efficiency of quantum circuits in feature extraction, paving the way for further research into the scalability of this approach. Specifically, the use of quantum circuits allows for richer transformations of input data, suggesting promising applications in more challenging datasets where classical machine learning models face limitations.

Future work will aim to refine the quantum circuit design, experiment with alternative quantum architectures, and test the model's generalizability on larger, more complex datasets. Additionally, exploring the integration of deeper quantum circuits and improved optimization strategies could further enhance the model's robustness and performance. We believe this research contributes to the evolving field of quantum machine learning, offering a stepping stone towards the development of more advanced quantum algorithms that leverage the unique computational advantages of quantum systems.

\end{document}